\documentclass{statsoc}
\usepackage[a4paper]{geometry}

\usepackage[utf8]{inputenc}
\usepackage{lipsum}

\usepackage{etoolbox}

\makeatletter
\patchcmd{\@makecaption}
  {\parbox}
  {\advance\@tempdima-\fontdimen2} 
  {}{}
\makeatother  

\usepackage[caption = false]{subfig} 
\usepackage{enumerate}
\usepackage{comment}
\usepackage{float}

\usepackage{graphicx}
\usepackage{tikz}
\usepackage{pgf}
\usepackage{pgflibraryarrows}
\usepackage{amsmath}
\usepackage{amsfonts}
\usepackage{amssymb}
\usepackage{url}
\DeclareSymbolFont{bbold}{U}{bbold}{m}{n}
\DeclareSymbolFontAlphabet{\mathbbold}{bbold}

\title[Discussion Laber et al.]
{Discussion of the paper ``Optimal treatment allocations in space and time for on-line control of an emerging infectious disease'' by E. B. Laber, N. J. Meyer, B. J. Reich, K. Pacifici, J. A. Collazo and J. Drake\thanks{This is the pre-peer reviewed version of the following article: Koskinen, J.H. (2018), Discussion of ``Optimal treatment allocations in space and time for on-line control of an emerging infectious disease'' by Laber, N. J. Meyer, B. J. Reich, K. Pacifici, J. A. Collazo and J. Drake, Journal of the Royal Statistical Society Series C, 67, p779, which has been published in final form at  https://rss.onlinelibrary.wiley.com/doi/full/10.1111/rssc.12266} }
\author[Koskinen]{Johan Koskinen}
\address{University of Manchester}

\begin{document}
\maketitle
\begin{abstract}
	This is a written discussion of the paper ``Optimal treatment allocations in space and time for on-line control of an emerging infectious disease'' by E. B. Laber, N. J. Meyer, B. J. Reich, K. Pacifici, J. A. Collazo and J. Drake, contributed to the Journal of the Royal Statistical Society Series C.
\end{abstract}	

\section{Disease spread on social networks}
This very thorough and welcomed contribution demonstrates the use of simulation in modern inference, and engages with several academic traditions, one of which, social network analysis, I would like to add some reflections on. Spread on networks has a long history in social network analysis (Coleman et al., 1957) and, particularly in the health science, there has been much recent attention paid to interventions (Valente, 2012; Morris, 2004). Estimating these processes has prompted the development of statistical models (such as Greenan, 2015) and simulation routines (e.g. Jenness et al., 2016). Rolls et al. (2012) investigate the behaviour of hepatitis C transmission on an empirical network (Aitken et al., 2008) and demonstrate effects of different types of network topologies on disease outcomes. Information on these disease-relevant networks is by necessity patchy but can be learned through sampling techniques such as in Rolls et al (2013a) or Handcock and Gile (2010). In fact, key features can to some extent even be estimated from respondent interviews (Krivitsky and Morris, 2017). Many of these methods rely on exponential random-graph models (Lusher et al., 2013), a class of log-linear models that model interactions between links that correspond to features such as the prevalence of hubs and the tendency towards clustering of ties. Figure 1 illustrates potential consequences of such features in a ‘realistic’ network (b) relative to an ‘unrealistic’ network (a) with the same number of links. The local clustering of network (b), means that links, rather than carrying the disease further from the seed node (black), out into the network, are used up linking to nodes that are already connected. Real life networks – both human and animal (even bats, see e.g. Willis, \& Brigham, 2004) – typically demonstrate high clustering (not merely artefacts of geography, Daraganova et al., 2012) and relatively short path-lengths. The example network N3 is meant to have these features (Robins et al., 2005) but to the naked eye appears to have rather long pathways. In fact the results, pairwise for S1 and N1, S2 and N2, and S3 and N3, respectively, seem to reflect the fact that the network topologies are more stringent forms of linkage than their spatial equivalents, resulting in similar behaviour but with lower uncertainty. How would the spread and interventions be affected by the more distinct network features? Obviously, these are big questions that have many moving parts (Rolls et al., 2013b) but the proposed allocation scheme is also very flexible.

\begin{figure}
\centering
    \makebox{
\includegraphics[scale=.5]{clustering.pdf}}
\caption{\label{fig:clustering}}
\end{figure}

\makeatletter
\renewcommand\@biblabel[1]{}
\makeatother

\end{document}